 \newcommand{\dm}{{\Delta_{max}}} 

\documentclass[draft]{agujournal2019}
\usepackage{url} 
\usepackage{soul}

\usepackage{xcolor}

\journalname{JGR: Space Physics}

\begin{document}

%
%

\title{Astrophysical explosions revisited: collisionless coupling of debris to magnetized plasma}

%
%




\authors{Ari Le\affil{1},
Dan Winske\affil{1},
Adam Stanier\affil{1},
William Daughton\affil{1},
Misa Cowee\affil{1},
Blake Wetherton\affil{1},
Fan Guo\affil{1}
}

\affiliation{1}{Los Alamos National Laboratory, Los Alamos, NM, USA}






\correspondingauthor{Ari Le}{arile@lanl.gov}




\begin{keypoints}
\item With hybrid particle-in-cell simulations, we formulate scalings for the size of an astrophysical explosion over a range of parameters.
\item Transport across the magnetic field by flute modes is quantified.
\item Estimates are developed for the fraction of the total debris that exits the explosion in free-streaming beams along the magnetic field.
\end{keypoints}

%
%

%
%


\begin{abstract}
The coupling between a rapidly expanding cloud of ionized debris and an ambient magnetized plasma is revisited with a hybrid (kinetic ion/fluid electron) simulation code that allows a study over a wide range of plasma parameters. Over a specified range of hypothetical conditions, simple scaling laws in terms of the total debris mass and explosion speed are derived and verified for the maximal size of the debris cloud and the fraction of debris that free-streams from the burst along the magnetic field. The amount of debris that escapes from the burst with minimal coupling to the background magnetic field increases with the debris gyroradius. Test cases with two different debris species--including a heavy minority species with a relatively large gyroradius--highlight how the collisionless coupling of the debris depends on the single-particle trajectories as well as the overall conservation of energy and momentum.
\end{abstract}

\section*{Plain Language Summary}
An astrophysical explosion is the rapid expansion of a cloud of ionized debris into a surrounding magnetized plasma. Examples include supernova remnants, targets hit by lasers in laboratory experiments, and active experiments in space where barium or other material is rapidly released. We systematically study the dependence of the distance the debris travels on its initial speed and on the total quantity of exploded mass. We also formulate simple estimates for the fraction of the debris that escapes along the magnetic field without losing much energy.

%
%

%


%
%
%
%

\section{Introduction}

Astrophysical explosions occur on a very wide range of spatial scales and require a diverse set of computer simulation techniques to model them. On very large scales, the explosion of massive stars leads to the collapse of the core, forming a black hole \cite{oconnor:2011}. Fluid models that include relativistic and gravitational effects are needed to correctly capture the dynamics. Here in our solar system, but still on very impressively large scales, coronal mass ejections and solar flares occur on the sun that are both visually stunning and can lead to space weather events on Earth. Magneto-hydrodynamic (MHD) simulations are necessary to model the global magnetic reconnection processes that cause the dramatic release of the sun’s energy into the solar wind and its propagation to the Earth \cite{gibson:1998}.  Closer to Earth, there are much smaller events that exhibit interesting plasma physics phenomena, such as the magnetic cavity (coma) in front of the nucleus of a comet that forms by the ejection of solar heated material into the solar wind and man-made, so-called active experiments in space, such as barium gas releases or nuclear explosions in the atmosphere \cite{ip:1987,bernhardt:1987,huba:1992,dyal:2006,goetz:2016}. 

In the space physics cases, the smaller spatial and temporal scales of the problem require a kinetic treatment of the ions (and sometimes the electrons as well). Winske and Gary \cite{winske:2007} performed numerical simulations to study such small explosions, with particular focus on the collisionless ion dynamics. That work provides motivation for the study here. Advances in computer hardware, algorithms, and physical understanding, as well as renewed interest in this subject over the past 15 years motivate the present work. For example, modern computer architectures can accommodate kinetic plasma simulations involving trillions of particles and billions of cells \cite{byna:2012} using codes that that utilize the computer resources extremely efficiently \cite{bowers:2008}. Recent laboratory experiments have investigated the early-time dynamics of the explosion process, using lasers to provide the energy source for the explosion and new diagnostics to investigate the interaction of the expanding plasma with the background plasma and magnetic field \cite{niemann:2014,schaeffer:2017,schaeffer:2018}. The larger simulations permit better modeling of the plasma dynamics \cite{clark:2013} and more sophisticated diagnostics provide better insight into the physics \cite{schaeffer:2014,schaeffer:2019}. 

The original calculations in Winske and Gary \cite{winske:2007} involved the expansion of a dense “debris” plasma expanding into a uniform magnetized background plasma. In those two-dimensional calculations the background magnetic field was in the plane of the simulation so that the dynamics involved both debris ion expansion across and along the magnetic field. The expanding debris ions interact with the background ions as well as the magnetic field, forming a magnetic cavity as the debris ions slow and eventually stop in the direction transverse to the magnetic field. The calculations were run for only a short time beyond the time of maximum cavity size. In the present study, the calculations can be run much longer in time and in larger domains to see how the cavity collapses and how debris ions propagate along and across the magnetic field during this process. In addition, the simulations are run in various geometries: in the two spatial dimensions with the magnetic field either in or perpendicular to the plane of the simulation. Also, the parameters of the simulations are varied to investigate scaling of quantities of interest. The simulations employ a kinetic hybrid model--where the debris and background plasma ions are treated as discrete particles and the electrons are considered as an inertia-less fluid \cite{winske:2003}. 

Even in a simplified uniform magnetized background plasma, this process is quite complicated. Here, the size of the magnetic cavity that is produced in these geometries is characterized as a function of a geometric factor, the so-called equal mass radius, and the initial expansion velocity of the debris ions, expressed in terms of the Alfven speed. How the size of the cavity and the furthest extent of the debris ions beyond the boundary of the cavity are modified when the initial debris mass is shared between a majority of light debris ions and a small fraction of heavier ions is also considered. We cover a wide range of parameters that spans from small systems where single-particle trajectories dominate the debris dispersal to large systems approaching MHD scales where bulk energy and momentum conservation yield estimates for the diamagnetic cavity size. We formulate new simple semi-empirical approximations for the distance traveled by the debris and the fraction of debris that free-streams away from the explosion along the magnetic field, covering the parameter space occupied by typical space and laboratory experiments.

\section{Diamagnetic Cavities}
First, we briefly provide a background on diamagnetic cavity formation, focusing on the conventions we adopt [primarily from \cite{winske:2007}]. Consider the expansion of a debris cloud into a uniform background plasma of density $n_b$ and magnetic field $B_0$. The background ions have mass $m_b$ and charge state $Z_b$. In the following, we normalize lengths to the background ion inertial length  $d_i = \sqrt{c^2\epsilon_0m_b/e^2Z_b^2n_b}$ and times to the inverse ion gyrofrequency $1/\omega_{ci} = m_b/eZ_bB_0$.

 A cloud of ionized debris expands rapidly into the uniform ambient background plasma. We consider in the following sections two cases: clouds composed of $N_d$ total ions all of mass $m_d$ and charge $Z_d$; and, to further understand the effects of different mass/charge states \cite{clark:2013}, clouds that include an additional heavier minority species of mass $m_h>m_d$. The cloud expansion is characterized by a radial velocity $V_d$, which may be normalized to the background Alfven speed $v_A = d_i\omega_{ci}$ to give an Alfven Mach number $M_A = V_d/v_A$. The expanding cloud generates a diamagnetic cavity \cite{winske:2019} by expelling plasma and magnetic field. Qualitatively for certain parameter regimes, the process occurs through ``Larmor'' coupling whereby the expanding cloud advects magnetic field lines (which are frozen into the electron fluid to a very good approximation) radially outwards. Ambient plasma is then accelerated by the induced electric field \cite{golubev:1978} and is transported with the field lines, forming a diamagnetic cavity. The cavity eventually collapses back down under the external magnetic and particle pressure as energy is radiated away, possibly in the form of a shock \cite{niemann:2014,schaeffer:2014}. The effectiveness of this coupling and its precise mechanisms depend on the plasma parameters.

There are two normalized measures of the total debris mass that have been used in previous analyses \cite{bashurin:1983,winske:2007,clark:2013}, and the relevance of each depends mainly on the Alfven Mach number $M_A$ of the explosion. A simple estimate of the maximal size of the debris cloud is found by assuming the initial kinetic energy of the debris $N_dm_dV_d^2/2$ is transferred to the kinetic energy of a volume of the background plasma accelerated to a speed on the order of $V_d$ and work done expelling the magnetic field from the diamagnetic cavity [see, e.g., \cite{clark:2013}]:
\begin{equation}
    \frac{1}{2}N_dm_dV_d^2 \approx \frac{1}{2}(n_bm_bV_d^2 + B_0^2/\mu_0)\times Vol^D(R)
    \label{eq:rest}
\end{equation}
Here, $Vol^D(R) = \pi R^2$ for 2D cylindrical geometries (and $Vol^D(R) = (4\pi/3) R^3$ for 3D spherical geometries). For large Alfven Mach numbers $M_A>1$, the debris is stopped mainly by the mass of background ions swept up in the outward explosion, and the first term on the right-hand side of Eq.~\ref{eq:rest} dominates. The debris then reaches approximately the so-called equal mass radius $\sim R_m$, defined as the radius that contains a mass of ambient plasma equal to the total debris mass: $N_dm_d = (n_bm_b)Vol^D(R_m)$. For sub-Alfvenic expansions ($M_A <1$), on the other hand, the magnetic field pressure plays a dominant role in slowing the cavity expansion. In this case, the pertinent length scale is the magnetic confinement radius $R_B$ given by balancing the explosion energy with the second term on the right-hand side of \ref{eq:rest}. This gives a magnetic confinement radius $R_B$ simply related to the equal mass radius by $R_B/R_m = M_A^{2/D}$. For a given expansion speed, the total normalized debris mass may be specified by either the equal mass radius $R_m$ (which we use here) or the magnetic confinement radius $R_B$.

\section{Hybrid Simulations}
\begin{figure}
\begin{center}
\includegraphics[width=0.8\textwidth]{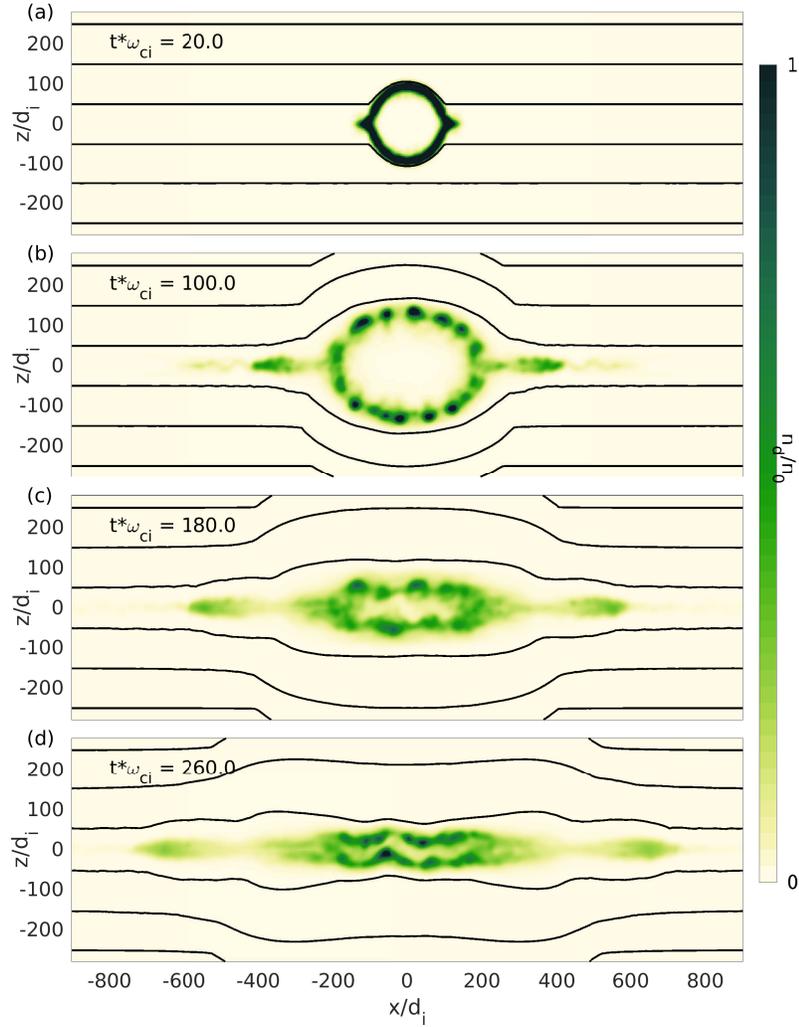}
\caption{Debris ion density at four different times over the course of a 2D hybrid simulation with initial magnetic field in the simulation plane. In-plane projections of magnetic field lines are drawn as black curves. This simulation used the same parameters as a previous study \cite{winske:2007}, with a debris ion mass of $m_d/m_b=3$, an initial expansion speed of $V_d=5v_A$, and an equal mass radius of $R_m = 150~d_i$.}
\label{fig:nd}
\end{center}
\end{figure}

To study the collisionless coupling of an exploding ionized debris cloud to a background magnetized plasma, we employ a hybrid particle-in-cell (PIC) code. The hybrid method combines a kinetic treatment of the ions with a simplified fluid model of the electrons \cite{winske:2003}. Here, our electron model reduces to an Ohm's law for the electric field of the form:
\begin{equation}
{\bf{E}} = -{\bf{u_i\times B}} - \frac{1}{ne}\nabla p_e +  \frac{1}{ne}{\bf{J \times B}} - \eta{\bf{J}} + \eta_H \nabla^2{\bf{J}}
\end{equation}
where quasi-neutrality is assumed so that $n=n_e=\sum_sZ_sn_s$ (including a sum over species $s$ of ions); the ion velocity appearing in Ohm's law is the charge-weighted ion current ${\bf{u_i}}=\sum_s Z_sn_s{\bf{u_s}}/n_e$; the current density is taken in the low-frequency approximation $\mu_0{\bf{J}} = \nabla\times{\bf{B}}$; and the normalized resistivity $\eta/(B_0/n_be)$ and hyper-resistivity $\eta_H/(B_0/n_bed_i^2)$ are set to small normalized values in the range of 0.001 to .005, which do not substantially affect the results of these simulations and are in line with previous hybrid simulations that agree with fully kinetic modeling \cite{le:2016,le:2018,wetherton:2019}. The electron pressure gradient accounts for electrostatic fields \cite{bonde:2015}, and the electrons are treated as an isothermal fluid. Test cases using an adiabatic electron equation of state ($p_e\propto n^\gamma$ with $\gamma=5/3$) showed little difference in the bulk dynamics of our simulations, which is primarily controlled by the electromagnetic coupling between the debris and background ions. Similar hybrid models have been used previously to study general astrophysical explosions \cite{bashurin:1983,winske:2007,brecht:2009,hewett:2011} and shocks \cite{thomas:1986,lembege:2001}, as well as problems related to chemical releases in the magnetosphere \cite{bernhardt:1987,delamere:1999} and laser-driven laboratory experiments \cite{clark:2013,weidl:2016,heuer:2020}.

Our hybrid code, Hybrid-VPIC, is built on the framework of the high-performance PIC code VPIC \cite{bowers:2008}, and it scales to the largest present computers. While the code may be run in 1, 2, or 3 spatial dimensions, we focus here on 2D simulations, which include both parallel and perpendicular plasma dynamics yet are small enough to allow a wide parameter scan. Hybrid-VPIC uses a standard explicit time advance algorithm including a 3rd- or 4th-order Runge-Kutta advance for the magnetic field, a leap-frog (Boris) time advance for the particles, and the simple linear extrapolation method \cite{winske:2003,karimabadi:2004} to compute ion currents for the time-advanced electric field. We have also performed cross-code verification of a subset of simulation results presented here against a separate hybrid code with a different numerical implementation \cite{stanier:2019}. We use nearest-grid-point particle weighting because it does not suffer from a non-physical numerical dispersion that can occur in hybrid codes irrespective of the time-stepping algorithm, and which can cause numerical inaccuracies in the structure of shock fronts when the ion inertial length is poorly resolved \cite{stanier:2020}. 

In a series of Hybrid-VPIC simulations, we vary the plasma parameters of the expanding cloud of ionized debris. In each case, the initial conditions contain a uniform background plasma of density $n_b$ and magnetic field of strength $B_0$. The electron and ion temperatures are both set to a value $T$ that gives $\beta_i = \beta_e = 0.1$ (where the background $\beta = \mu_0n_bT/B_0^2$ for each species). The grid spacing resolves at least the ion inertial length with $dx=d_i$ (and $dx=0.5d_i$ for the smaller simulations), and the time step is $dt=0.01/\omega_{ci}$. Most of the simulations presented here used periodic boundary conditions with the domain size large enough (ranging from $\sim500~d_i$ to $\sim7200~d_i$ in the magnetic field-aligned direction) so that re-circulating particles do not interact with the diamagnetic cavity. The grid sizes and resolutions are listed in Appendix A. A few additional cases were run with open boundary conditions \cite{daughton:2006} that absorb outgoing particles, and the main results were unchanged. The background plasma ions are represented by $200$ numerical particles per cell, requiring $\sim3$ billion numerical particles total for the largest simulations. 

An exploding cloud of ionized debris is introduced at the center of the simulation domain. We performed a set of simulations spanning the ranges of expansion Alfven Mach numbers $M_A$ = $V_d/V_A$ = 0.5, 1, 2, 5, 10 and equal mass radii of $R_m=$ 10~$d_i$, 40~$d_i$, and 150~$d_i$. Similar to a previous study \cite{winske:2007}, we choose debris ions of mass $m_d = 3m_b$ (where $m_b$ is the mass of the background ions). The charge states are equal, $Z_d = Z_b$. Additional simulations were performed with an added minority species of ions with a heavier mass $m_h = 5m_d = 15m_b$ (and $Z_h = Z_b$). The mass of $m_h=15m_b$ was selected so that the simulations would span a range with debris gyroradii both smaller and larger than the equal mass radius $R_m$ to gain additional insight on the effect of the ion species mass on transport. The main debris ion population cloud is initialized with a Gaussian density profile $n_{d} = n_{d0}\exp(-R^2/L^2)$,
where the length scale $L$ depends on $R_m$ and is taken as $L=0.1R_m$. In the runs with an additional minority heavy species, the heavy species is loaded with a similar profile, $n_{h} = n_{h0}\exp(-R^2/L^2)$, where we use a relative fraction of heavy ions of $f_h = n_{h0}/n_{d0}$. The peak density $n_{d0}$ is determined by the equal mass radius $R_m$. It is given explicitly by $n_{d0}/n_b = (R_m^2/L^2)(m_b/m_\ast)$, where $m_* = m_d + f_hm_h$. The initial debris ion velocity distribution is a shifted Maxwellian with the mean flow directed radially outward at a speed $V_d$ and with a thermal spread given by $\sqrt{T_d/m_d} = 0.1V_d$. When we refer to the debris gyroradius below, it is with respect to the initial expansion speed: $\rho_d = V_d/\omega_{ci}$. Each species of debris ions is sampled with a number of numerical particles equal to $1/50$th the number of numerical background particles, implying a range of 1 to 60 million numerical debris particles. The numerical weights of these PIC macroparticles are different from those of the background particles, and they are set so that the physical densities correspond to the peak values given by $n_{d0}$ and $n_{h0}$.

Sample snapshots at four times over the course of a typical simulation are plotted in Fig.~\ref{fig:nd}, where the parameters $R_m=150~d_i$ and $M_A=5$ are identical to a simulation presented in \cite{winske:2007}. The plots show contours of debris density in green along with sample in-plane magnetic field lines over-plotted as black curves. At early time in Fig.~\ref{fig:nd}(a), the cloud is still rapidly expanding at nearly its initial velocity. In Fig.~\ref{fig:nd}(b), the cloud of debris has reached its largest radial size. Beams of parallel-streaming ions escaping in the $\pm x$ directions also become apparent. The magnetic field even away from the debris is perturbed by a magnetosonic shock that is launched by the expansion \cite{heuer:2020}. Figs.~\ref{fig:nd}(c-d) show the debris cloud during the collapse phase as the ambient magnetic field and plasma pressure cause the cavity to deflate.

\section{Perpendicular Cavity Size with Field in the Plane}
\label{sec:perp}

A basic question about the dynamics of astrophysical explosions is how far the debris travels from the initial blast point. Details of the streaming ions and expanding shock waves will be treated elsewhere. While analytical models are possible for the initial phases of coupling of the debris to the background plasma \cite{golubev:1978, bashurin:1983}, the longer-time nonlinear phases require numerical simulation \cite{winske:2007, hewett:2011}. Nevertheless, simple estimates give a good approximation for the maximal size of the debris cavity, and small empirical corrections can be applied based on the simulations.

\begin{figure}
\begin{center}
\includegraphics[width=0.75\textwidth]{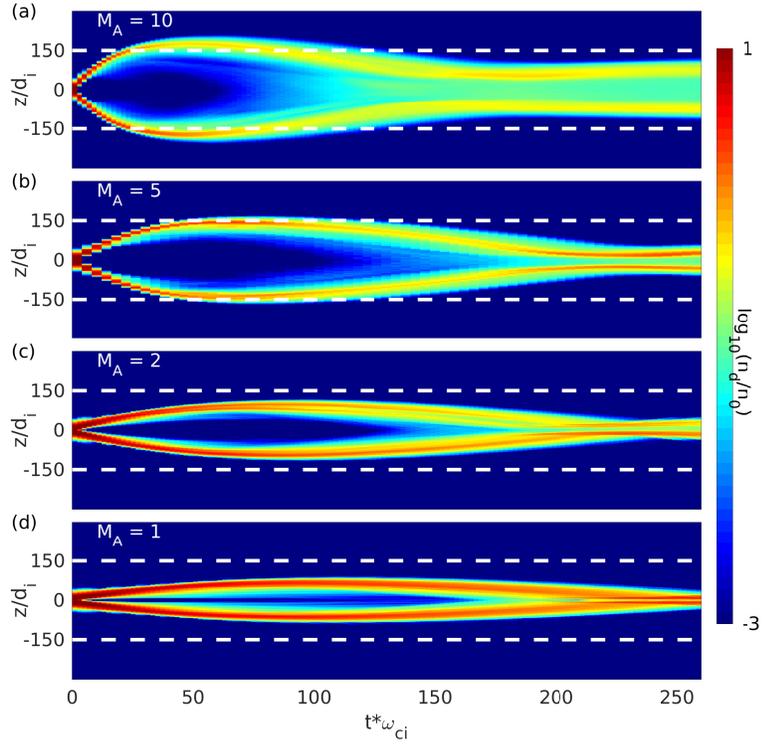}
\caption{Density along a vertical cut at $x=0$ over time in a set of simulations with an equal mass radius of $R_m=150d_i$ (marked as dashed horizontal while lines) similar to Fig.~\ref{fig:nd}, with initial debris expansion Mach numbers $M_A=V_d/v_A$ of (a) 10, (b) 5, (c) 2, and (d) 1.}
\label{fig:zcut}
\end{center}
\end{figure}

We adopt a metric for the perpendicular cavity size used by \cite{winske:2007}, a metric that is based on the debris density rather than the magnetic field profile. The definition of the size of the debris cloud over time $\Delta_{98}(t)$ is based on the debris density profile along the vertical cut at $x=0$ perpendicular to the background magnetic field. It is defined implicitly by
\begin{equation}
    \int_0^{\Delta_{98}} n_d(t,x=0,z) dz = 0.98 * \int_0^\infty n_d(t,x=0,z) dz,
    \label{eq:delta98}
\end{equation}
which thus defines $\Delta_{98}$ as the radial distance that contains $98\%$ of the mid-plane debris density. The maximal size of the debris cavity $\dm$ is then defined as the maximum over time of $\Delta_{98}(t)$. The debris density profile along the central vertical cut ($x=0$) is plotted over time from a series of simulations with $R_m=150~d_i$ in Fig.~\ref{fig:zcut}. As visible in Fig.~\ref{fig:zcut}, the time when the cloud reaches its maximal radial extent depends on the initial speed of the expansion. Typically, expansions with faster initial speeds reach their maximal size in shorter times, and the maximal extent of the debris increases with $V_d$. 

\begin{figure}
\begin{center}
\includegraphics[width=0.7\textwidth]{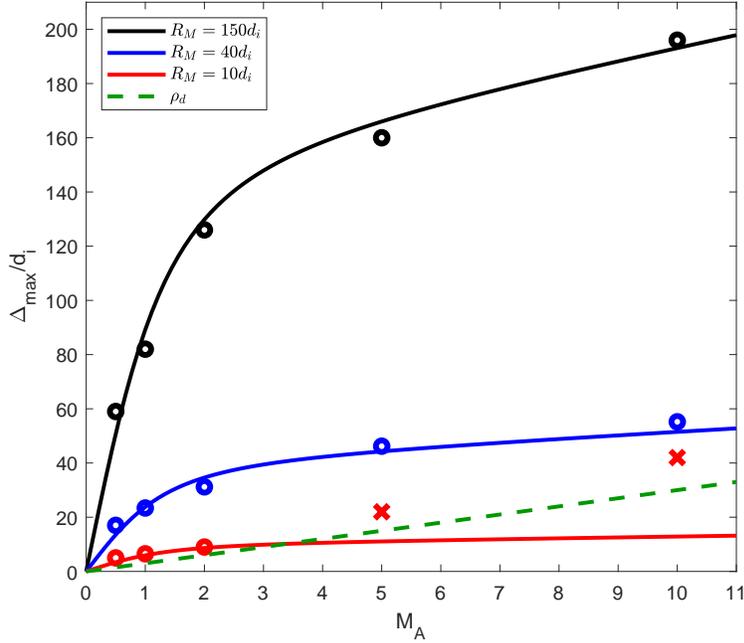}
\caption{The maximal perpendicular size $\dm$ of the debris cloud from a set of 2D hybrid simulations with the magnetic field in the simulation plane with varying initial expansion Alfven Mach numbers $M_A = V_d/v_A$ and equal mass radii $R_m$ of (black) 150 $d_i$, (blue) 40 $d_i$, and (red) 10 $d_i$. The curves are given by the fit in Eq.~\ref{eq:deltamax}. The two points marked with red $\times$s are in the ``decoupled'' coupled regime \cite{hewett:2011}, with the debris gyro-radius $\rho_d>R_m$.}
\label{fig:deltamax}
\end{center}
\end{figure}

Equation \ref{eq:rest} implies a simplified scaling in terms of the equal mass radius $R_m$ and expansion Alfven Mach number $M_A$ for maximal debris cloud size of $\dm\sim R_m*(1+1/M_A^2)^{-1/D}$. Based on the form of that scaling, we find an approximate interpolation fit for the normalized size $\delta_{max}=\dm/R_m$ of the debris clouds in our set of 2D simulations with magnetic field in the plane of the form:
\begin{equation}
    \delta_{max} = \frac{\dm}{R_m}\sim \frac{1 + c_0 M_A}{(1 + c_1/M_A^2)^{1/D}},
        \label{eq:deltamax}
\end{equation}
where for our 2D simulations $D=2$, and we find fitting parameters of $c_0=0.03$ and $c_1=2$ for the runs with $m_d/m_b=3$ and $Z_d/Z_b=1$. In a qualitative sense, the parameter $c_0$ may be thought of as a small correction accounting for the debris ion gyro-radius $\rho_d\propto V_d$ increasing at larger expansion $M_A$. While $c_0$ helps improve the fit to the data over the range of $M_A$ and $R_m $we considered, we do not expect it to be accurate for extrapolation. As described below, Eq.~\ref{eq:deltamax} only applies for systems with $\rho_d<R_m$, and we necessarily have $c_0\ll 1$. The parameter $c_1$, meanwhile, may be thought of as accounting for the fact that our simple estimate of the cloud size neglects the bending and compression of the initially straight magnetic field lines. Assuming the diamagnetic cavity is essentially devoid of magnetic field and surrounded by practically circular magnetic field lines, the net inward magnetic tension force (associated with a large-scale shear Alfven wave launched by curving the field line) will be the same size as the magnetic pressure force. This explains why $c_1$ is of order unity, and we expect it to be of similar magnitude for values of $R_m$ larger than those considered here. These fitting parameters depend weakly on the initial conditions, particularly the initial size of the debris cloud and the charge and mass states of the debris ions. The fitted scaling of Eq.~\ref{eq:deltamax} is plotted as solid curves along with simulation data in Fig.~\ref{fig:deltamax}. In addition, the size of the debris gyroradius, $\rho_d = V_d/\Omega_{cd}$, is shown as a function of $M_A$ as a dashed green line.

The majority of the simulations obey the simple scaling as in Eq.~\ref{eq:deltamax}. Two exceptions are the two fastest expansion cases ($M_A$ of 5 and 10) marked with red $\times$s in Fig.~\ref{fig:deltamax} for the simulations with an equal mass radius of $R_m=10~d_i$. In fact, it is expected that these two cases fall into a different regime where the motion of the debris ions is limited by single particle orbit effects rather than the collective formation of a diamagnetic cavity. This ``decoupled'' regime \cite{hewett:2011} develops in terms of our simulation parameters roughly when $\rho_d > R_m$, when the gyroradius of individual debris ions becomes large compared to the nominal diamagnetic cavity size.

To further examine the effects of a large debris ion gyroradius, we performed a set of simulations identical to those above but with the addition of a heavier minority ion species of mass $m_h = 5m_d = 15m_b$. The heavy species density is initially set as $n_h = 0.05n_d$. Meanwhile, the total mass of the debris cloud, including both species of masses $m_d$ and $m_h$, is held fixed to give the same equal mass radii $R_m=$ 10, 40, and 150~$d_i$  as in the first set of simulations. We scan the Alfven Mach number over the values 1, 2, 5 and 10. 

\begin{figure}
\begin{center}
\includegraphics[width=1.0\textwidth]{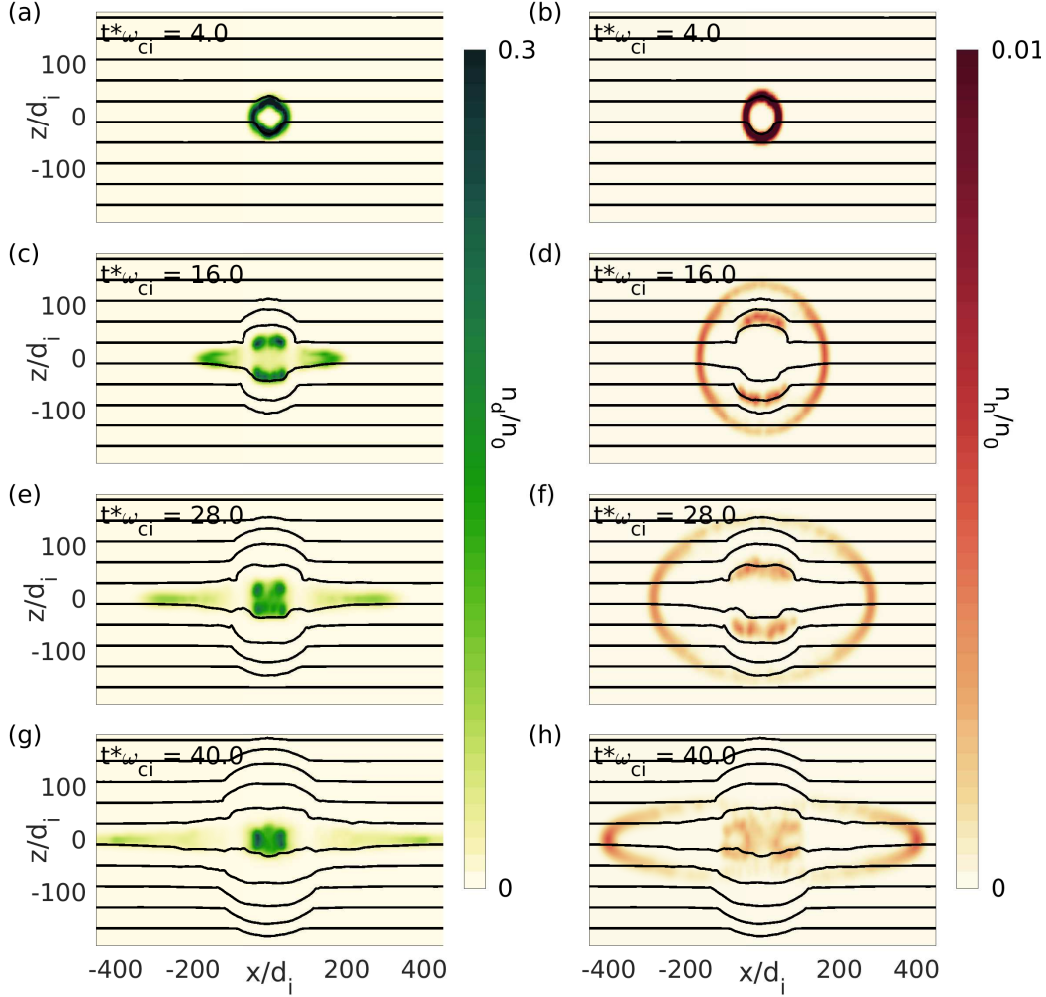}
\caption{Density of (left) main debris ion species with $m_d=3m_b$ and (right) minority heavier species with $m_h=15m_b$ at four different times over the course of a 2D hybrid simulation. In-plane magnetic field lines are drawn as black curves. This simulation has an equal mass radius of $R_m=40~d_i$ and an expansion Alfven Mach number of $M_A=10$.}
\label{fig:nd-h}
\end{center}
\end{figure}

Example density profiles of the two debris ion species are plotted in Fig.~\ref{fig:nd-h} from a calculation with an equal mass radius of $R_m=40~d_i$ and an expansion Alfven Mach number of $M_A=10$. We select this case because the lighter debris ions have a nominal gyroradius of $\rho_d=30~d_i < R_m$, while the heavier minority species has a gyroradius of $\rho_h=150~d_i > R_m$. As seen in Fig.~\ref{fig:nd-h}(c), the majority lighter ions reach approximately the equal mass radius before the diamagnetic cavity collapses back down. The heavier ions, on the other hand, split into two main populations as in Fig.~\ref{fig:nd-h}(d). A portion of the heavy debris couples to the electric fields in the cavity and only reach marginally farther than $R_m$ from the initial blast point. The rest of the ions form a low-density ``de-coupled'' \cite{hewett:2011} population that essentially undergoes unimpeded gyro-motion about the ambient magnetic field. This population reaches $\sim\rho_h\sim150~d_i$ from the initial burst before gyrating back inwards.

\begin{figure}
\begin{center}
\includegraphics[width=0.7\textwidth]{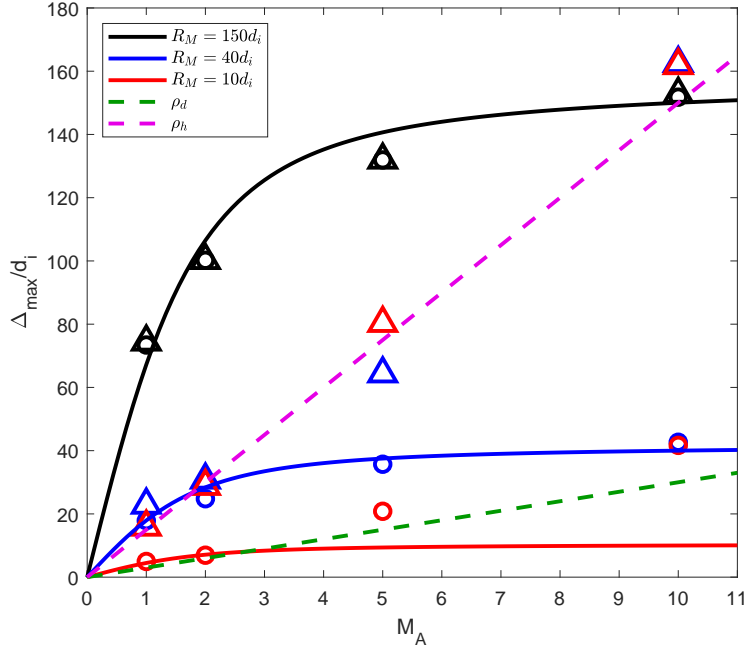}
\caption{Similar to Fig.~\ref{fig:deltamax}, the maximal perpendicular size $\dm$ of the debris cloud from a set of 2D hybrid simulations with the magnetic field in the simulation plane. Two ion species are included here: a main species as before with $m_d = 3m_b$ and a heavier minority species with $m_h = 15m_b$ with a relative number density of $n_h/n_d = 0.05$.  The equal mass radii $R_m$ are (black) 150 $d_i$, (blue) 40 $d_i$, and (red) 10 $d_i$. The solid curves are given by the fit in Eq.~\ref{eq:deltamax} with $c_0 = 0.005$ and $c_1=4$. The circles are the main debris species and the triangles are the heavier debris species. The dashed lines give the nominal gyroradii $\rho_d$ (green) of the main debris species and $\rho_h$ (magenta) of the heavier minority species.}
\label{fig:deltamax-h}
\end{center}
\end{figure}

Figure~\ref{fig:deltamax-h} shows the maximal debris cloud size $\dm$, now defined separately for each debris species, from the set of simulations with the minority heavy species. The circles in the figure are the main debris ions, and the triangles are the heavier species. The dashed curves show the gyroadii $\rho_d$ and $\rho_h$ for each species, while the solid curves give an interpolation fit for the lighter debris species as in Eq.~\ref{eq:deltamax} with $c_0=0.005$ and $c_1=4$. For the largest runs with $R_m=150~d_i$, we have $\rho_d$ and $\rho_h \le R_m$. In all these cases, a simple estimate for $\dm$ agrees with the size of the expanded debris cloud for both species, and the two species lie practically on top of each other. For the smaller normalized clouds with $R_m=$ 40 and 10 and larger $M_A$, on the other hand, the heavy debris gyroradius is large with $\rho_h>R_m$. In these cases, a portion of the heavy ions decouple, and their motion is limited by single particle orbit effects. The lighter debris, meanwhile, continues to couple to the background field and plasma. Note that in the cases where the heavy species $\rho_h>R_m$, the heavy ions split into two populations as in Fig.~\ref{fig:nd-h}, with one confined to the diamagnetic cavity and the other undergoing nearly free gyromotion about the magnetic field. The measure $\dm$ is based on the total heavy debris density and reaches out to the freely gyrating population at maximal distance of $\rho_h$ for the cases where $\rho_h>R_m$.

A distinction should be made between the motion of debris away from the initial blast point and motion across the magnetic field. In the 2D $x-z$ plane simulations here with magnetic field in the simulation plane, the in-plane magnetic field lines may be described by contours of the vector potential component $A_y(x,z)$. For our simplified geometries, the vector potential may be expressed as the magnetic flux
\begin{equation}
    A_y(x,z,t) = -\int_{z_0}^z B_x(x,z,t) dz,
    \label{eq:ay}
\end{equation}
where $z_0$ is arbitrary as long as it remains in a region of unperturbed magnetic field. Example in-plane magnetic field lines are plotted in Fig.~\ref{fig:nd}.  Because of the imposed 2D symmetry, canonical $y$ momentum of each debris ion is conserved \cite{golubev:1978,bashurin:1983}, so that for each debris ion $P_{y0} = A_y + (m_d/eZ_d)v_y $ is constant. Following Eq.~\ref{eq:ay} and noting that magnetic flux is conserved to an excellent approximation, this sets a limit on the maximal distance $z$ a given ion may reach. As a result, the ions in 2D geometries are strictly tied to an initial magnetic flux surface defined by $A_y = A_{y0}$, and they can only travel a distance of order $\rho_{d} = v_y/\omega_{ci}\sim v_D/\omega_{ci}$ from this magnetic surface (which will nevertheless evolve over time as the diamagnetic cavity expands and then collapses). The perpendicular transport is therefore strictly limited in 2D simulations with in-plane magnetic field lines \cite{jokipii:1993}. In addition, as the magnetic cavity collapses, the ions will exactly follow the field lines as they return toward their initial configuration. In Section~\ref{sec:by}, we consider geometries with the magnetic field perpendicular to the simulation plane that do not share the same constraints on ion and magnetic field line motion.

\section{Free-streaming Ion Beams}
\label{sec:beam}

Debris ions moving mainly parallel to the background magnetic field couple weakly to the ambient plasma. Rather than coupling through the Larmor mechanism \cite{bashurin:1983,golubev:1978}, the parallel-streaming ions couple to the background primarily through ion-ion beam instabilities \cite{gary:1984,weidl:2019,heuer:2020}. This coupling is comparatively weak, and in all our cases, a population of ions streams nearly freely from the explosion for at least several hundred ion inertial lengths. The incipient beams are visible in the two topmost panels of Fig.~\ref{fig:nd}.

Based on an approximate theory of the very early-stage debris-background coupling, it was found that ions whose expansion velocity makes an angle less than $\theta_\ast\sim \sin^{-1}(1.4\rho_d/R_m)$ should be weakly coupled to the background magnetic field and plasma \cite{bashurin:1983}. The approximations for this estimate clearly break down when $\rho_d>R_m$, and in that case the ions are in the ``decoupled'' regime \cite{hewett:2011}. Qualitatively, the angle for strong coupling suggests that there will be a ``hole'' with an opening angle proportional to $\rho_d/R_m$ through which debris ions may stream away from the explosion. Based on simple geometric arguments (and for small $\theta_\ast$), a fraction $F\sim2\theta_\ast/\pi$ of the debris ions will stream away from the explosion along the magnetic field with weak coupling to the background.

In the simulations, we do not observe a hard cut-off in velocity and pitch angle with respect to the magnetic field for debris ions to stream relatively freely away from the explosion site. Rather, there is a diffuse tail of ions with a range of velocities ranging from the initial expansion speed $V_d$ down to a little over the background Alfven speed $V_A$. To quantify the fraction of debris ions in the parallel-streaming beams, we adopt the following measure: at the time when the debris cloud reaches its maximal size $\dm$, we find the fraction $F_{90}$ of the total debris ion mass contained in regions where the bulk debris ion velocity $|u_x|>0.90V_d$ ($x$ is in the direction of the ambient  magnetic field). The velocity threshold of $90\%$ of $V_d$ is somewhat arbitrary, though it gives a reasonable measure of the fraction of free-streaming debris ions. A range covering $80-95\%$ of $V_d$ gives approximate debris fractions that are within a factor of $\sim2$ of each other.

\begin{figure}
\begin{center}
\includegraphics[width=0.7\textwidth]{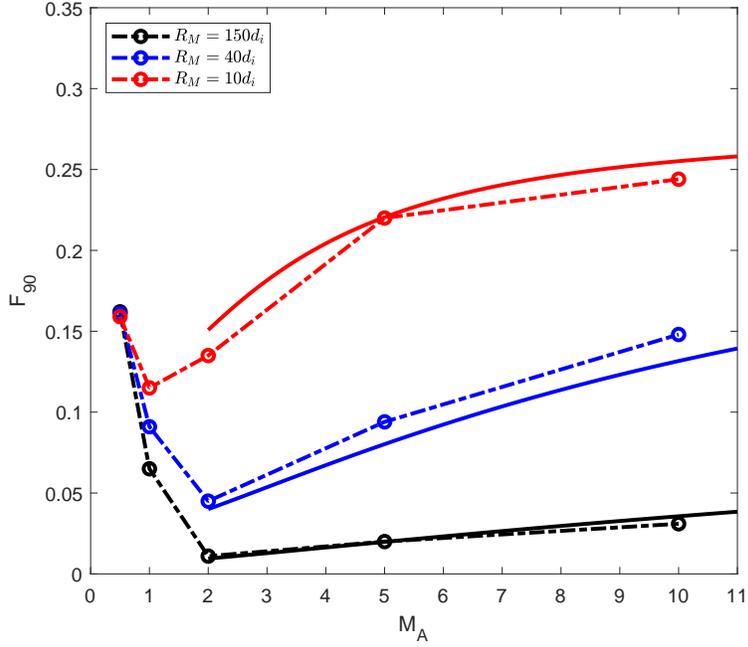}
\caption{The fraction $F_{90}$ of debris ions contained in the free-streaming parallel beams across the range of parameters used in this study. The beams are defined as regions where the debris ion bulk velocity $u_{dx}$ ($x$ is the direction of the background magnetic field) is $>90\%$ of the initial expansion speed $V_d$, and $F_{90}$ is measured at the time when the debris cavity size reaches its maximum value $\dm$. Dashed curves connect the data points to guide the eye. The solid curves are an approximate fit for $M_A>2$.}
\label{fig:f90}
\end{center}
\end{figure}

\begin{figure}
\begin{center}
\includegraphics[width=0.7\textwidth]{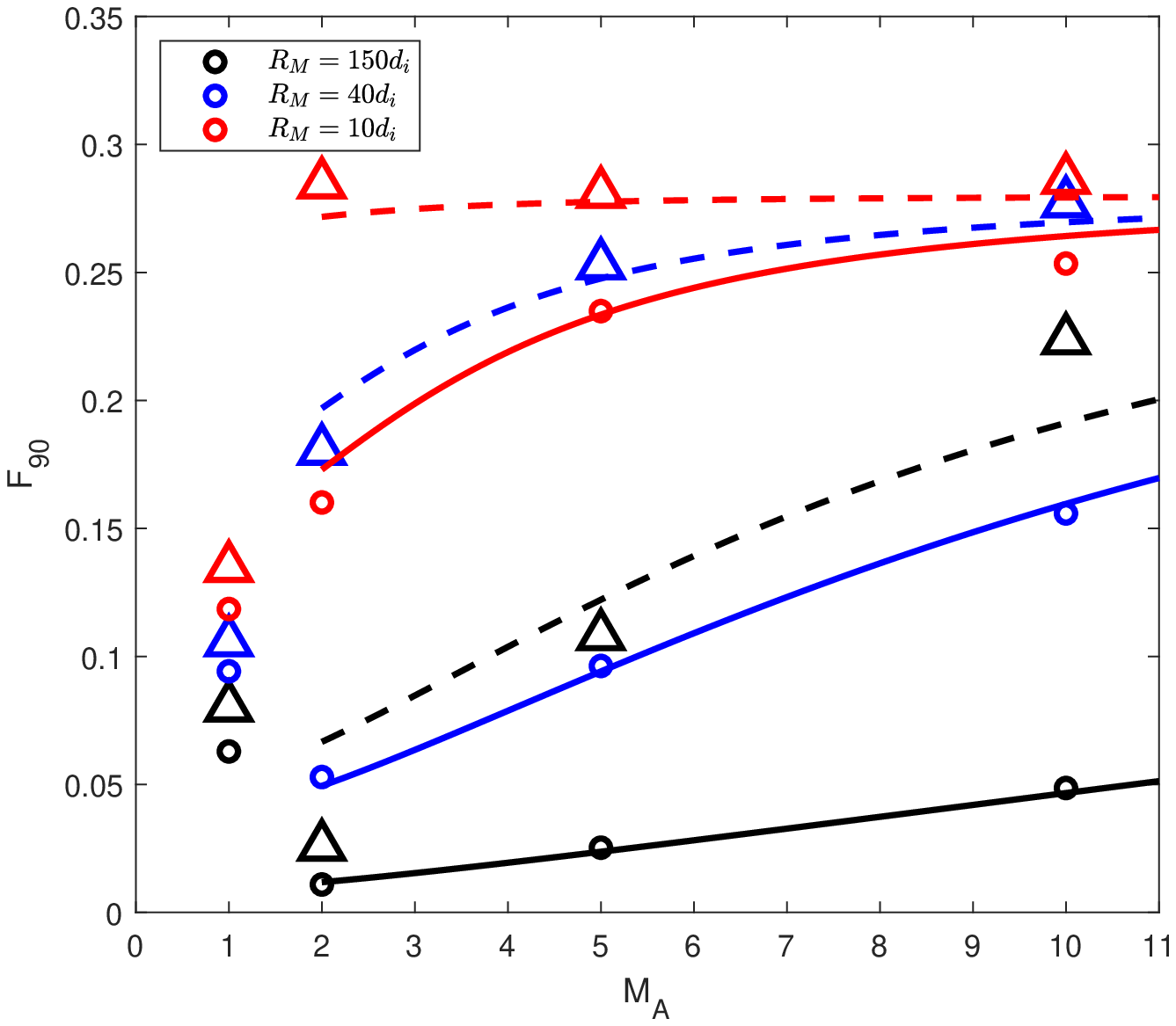}
\caption{Plots similar to Fig.~\ref{fig:f90} of the fraction $F_{90}$ of debris ions contained in the free-streaming parallel beams, now with an additional heavy species of mass $m_h=15m_b$. The curves give an approximate fit of $F_{90}$ based on the gyroradius of the main debris species (solid curves and circles) and the heavy minority species (dashed curves and triangles) following Eq.~\ref{eq:f90}. Circles mark data for the main debris species, and triangles mark the heavy minority debris species.}
\label{fig:f90-h}
\end{center}
\end{figure}

The measure $F_{90}$ of the fraction of free-streaming debris is plotted from a series of simulations in Fig.~\ref{fig:f90}. For sub-Alfvenic expansions, the fraction $F_{90}$ is a strong function of the expansion speed $M_A$, and it appears to be more weakly dependent on $R_m$ for $M_A<1$. For Alfven Mach numbers $M_A>2$, we find a rough fit of the simulation data for $F_{90}$ of the form 
\begin{equation}
F_{90}\sim C_{90}*\sigma(x)
\label{eq:f90}
\end{equation}
where for 2D cylindrical explosions, the asymptotic value $C_{90} = 2\arccos(0.9)/\pi\sim0.29$ (which corresponds to the initial fraction of debris at $t=0$ with $|v_x|>0.9V_d$), $\sigma(x)=(x/\sqrt{1+x^2})^{1.1}$, and $x=\rho_d/\dm$. This form is based on the considerations above that suggest the fraction should scale as $\propto \rho_d/R_m$ for small gyroradius $\rho_d$ and reach a plateau given by the initial velocity distribution as $\rho_d/R_m$ approaches 1. Figure~\ref{fig:f90-h} shows data similar to Fig.~\ref{fig:f90} for the simulations with the addition of a heavy ($m_h=5m_d=15m_b$) minority debris species. The free-streaming fractions of the main debris species are shown as circles with the solid lines based on Eq.~\ref{eq:f90} using $x=\rho_d/\dm$, and free-streaming fractions of the heavier debris species are shown as triangles with the dashed lines based on Eq.~\ref{eq:f90} using $x=\rho_h/\dm$. The heavier species generally has a larger fraction that escapes the explosion in fast beams directed along the magnetic field.

\section{Perpendicular Transport with Flute Modes}
\label{sec:by}

\begin{figure}
\begin{center}
\includegraphics[width=0.4\textwidth]{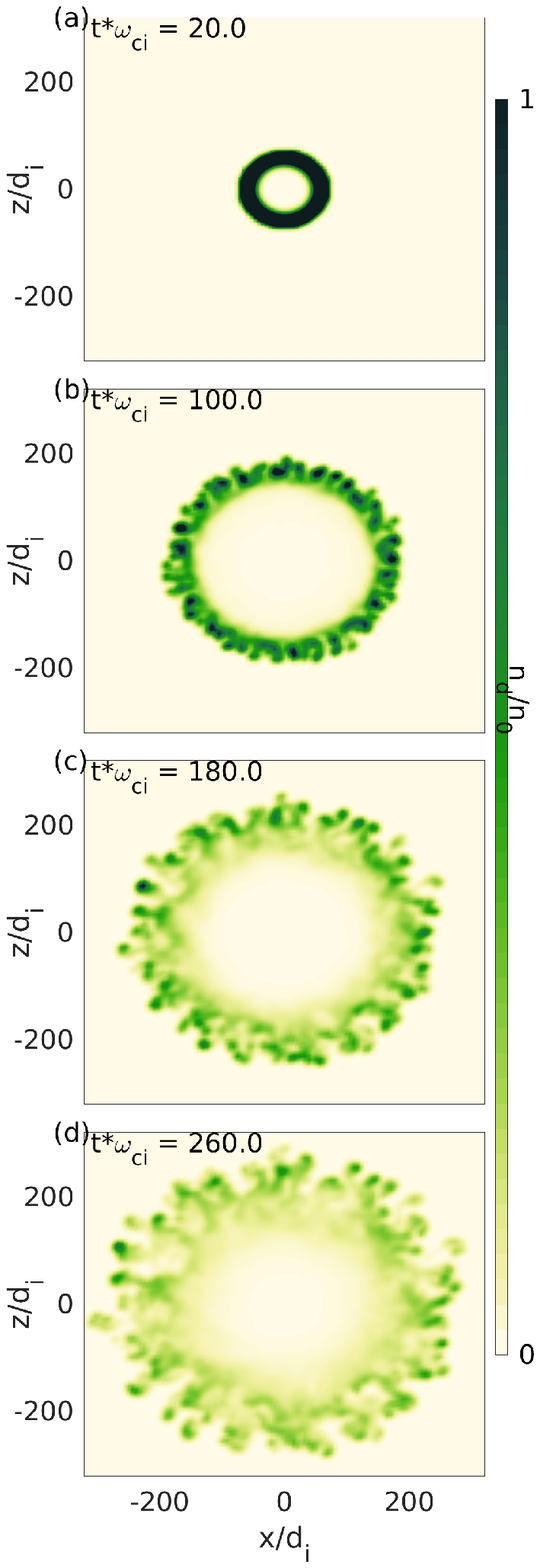}
\caption{Plots of debris ion density similar to Fig.~\ref{fig:nd}, but with the initial magnetic field into the simulation plane. The basic plasma parameters are identical to the simulation in Fig.~\ref{fig:nd}: $m_d/m_b=3$, $V_d=5v_A$, and $R_m = 150~d_i$. 2D simulations with the magnetic field perpendicular to the plane allow the development of flute modes as the debris decelerates against the background \cite{colgate:1965, hassam:1987,winske:1988,sgro:1989,winske:2019}.}
\label{fig:ndby}
\end{center}
\end{figure}

\begin{figure}
\begin{center}
\includegraphics[width=0.8\textwidth]{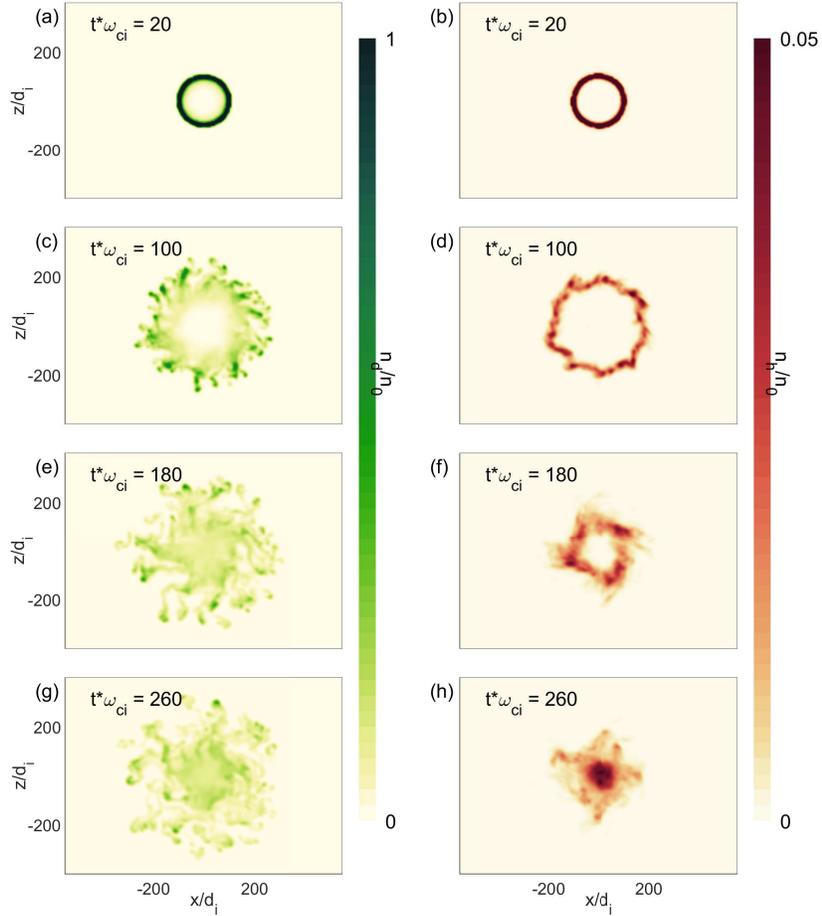}
\caption{Plots of debris ion density similar to Fig.~\ref{fig:ndby} for (left) the majority debris ion species and (right) a heavy ($m_h=5m_d=15m_b$) minority species.}
\label{fig:ndbyh}
\end{center}
\end{figure}

In this section, we perform 2D simulations similar to those of Section~\ref{sec:perp}, but now with the magnetic field perpendicular to the simulation plane. While this geometry does not track the parallel dispersion of the expanding plasma cloud, it captures flute-like interchange modes \cite{ripin:1987,colgate:1965,huba:1987,zakharov:2006,dyal:2006, winske:2019} that allow the transport of plasma across the magnetic field. The flute modes captured by the hybrid model are modified Rayleigh-Taylor modes driven by the deceleration of the debris cloud, and they do not include smaller-scale lower hybrid drift modes.

Data from same times and with the same nominal plasma parameters of $R_m=150~d_i$ and $M_A=5$ as the data in Fig.~\ref{fig:nd} are plotted for a simulation with the magnetic field into the simulation plane in Fig.~\ref{fig:ndby}. Again, the plots show contours of debris ion density. The flute modes that develop at the edge of the cavity now allow turbulent plasma motion that carries debris ions across the magnetic field. Importantly, the constraint on cross-field transport imposed by the 2D symmetry described at the end of Section ~\ref{sec:perp} does not hold for a magnetic field perpendicular to the plane. As a result, the debris ions are not strictly tied to the magnetic field lines, and the flute modes allow a genuine slipping of plasma across the magnetic field. 

A similar plot is shown for a simulation with an additional heavy species with $m_h=5m_d=15m_b$ and density $n_h=0.05n_d$ in Fig.~\ref{fig:ndbyh}. The majority debris species density is shown on the left, and the heavy species density is shown on  the right. A significant fraction of the heavy debris ions are ``decoupled,'' and they retain a large fraction of their initial energy. These heavy ions continue to undergo Larmor motion with a relatively large gyroradius ($\rho_h\sim 75~d_i$). The lighter ions couple to the background and are slowed down. The lighter ions are also the ones that are effectively transported across the magnetic field by the flute modes. The majority of the lighter ions therefore form a more diffuse cloud and do not gyrate back into the center of the magnetic cavity.

\begin{figure}
\begin{center}
\includegraphics[width=0.7\textwidth]{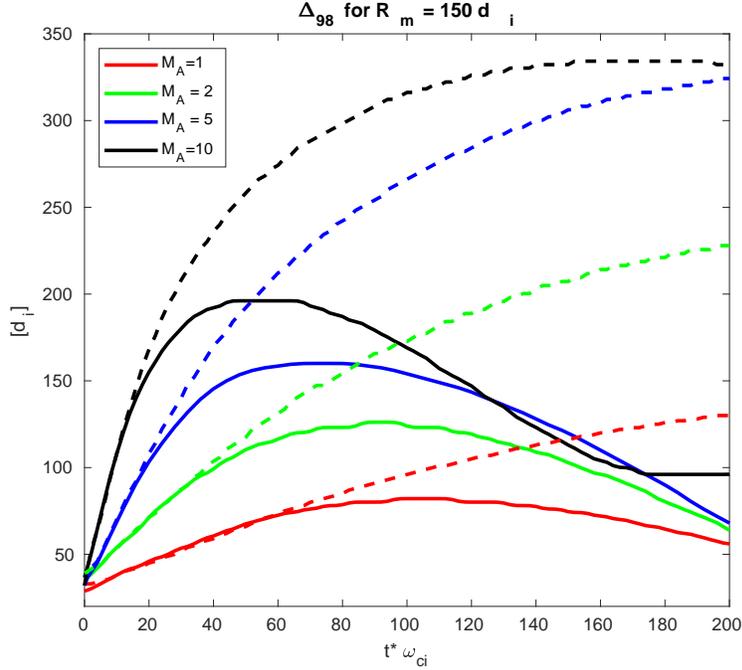}
\caption{The size $\Delta_{98}(t)$ of the debris cloud over time for runs with an equal mass radius of $R_m = 150~d_i$ and a single debris ion species. The size is defined as in Eq.~\ref{eq:delta98}. The solid curves are from simulations with the magnetic in the simulation plane, and the dashed curves are from simulations with the magnetic field perpendicular to the simulation plane. Flute instabilities carry debris ions a few gyroradii beyond the equal mass radius $R_m$.}
\label{fig:delta98}
\end{center}
\end{figure}

We compute the size of the debris clouds using the measure given by Eq.~\ref{eq:delta98}, using an average over all radii emanating from the burst point rather than the single vertical cut used for the previous orientation of the ambient magnetic field. In Fig.~\ref{fig:delta98}, we plot the size $\Delta_{98}$ of the cavities over time for simulations with $R_m=150~d_i$ both (solid curves) with magnetic field in the simulation and (dashed) with magnetic field perpendicular to the simulation plane. In each case, the dashed curves lie well above the solid curves, highlighting the efficiency of flute instabilities in transporting debris.

\begin{figure}
\begin{center}
\includegraphics[width=0.75\textwidth]{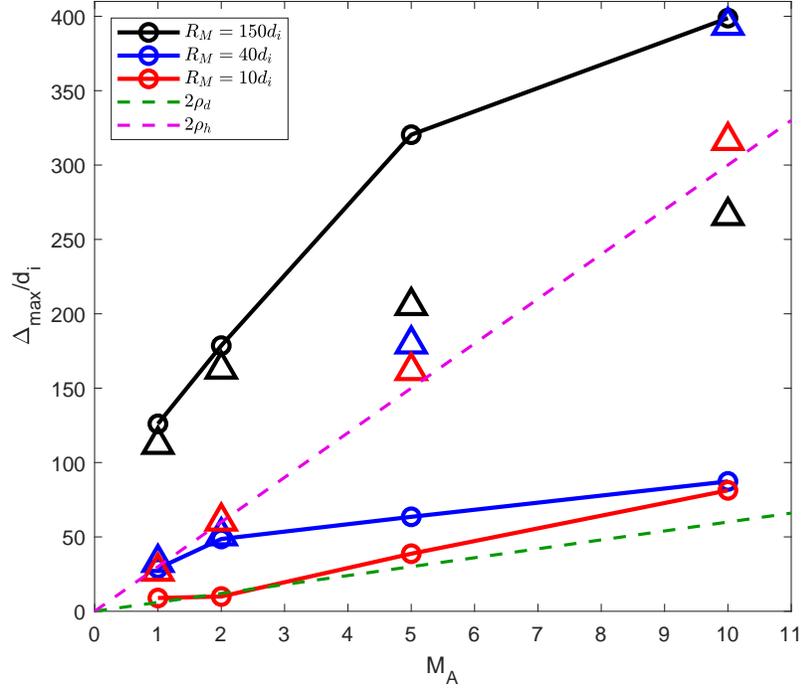}
\caption{The size maximal size of the debris clouds $\dm$ from simulations with two ion species and the magnetic field perpendicular to the simulation plane. Circles are the main debris species and triangles are the heavy minority species. The perpendicular transport of debris ions is enhanced by flute modes in each case. The dashed curves give twice the majority debris gyroradius $\rho_d$ (green) and heavier minority gyroradius $\rho_h$ (magenta), which set a lower bound based on single particle orbits. }
\label{fig:deltaby}
\end{center}
\end{figure}

In Fig.~\ref{fig:deltaby}, we plot the maximal size $\dm$ of the cavities with the magnetic field into the simulation plane. The data plotted here are for simulations that included the additional heavier minority debris species as in Section~\ref{sec:perp} and Fig.~\ref{fig:ndbyh}. Note that the solid lines are not fits to data, but are drawn simply connecting data points to guide the eye. Similar to the case for in-plane magnetic field in Fig.~\ref{fig:deltamax-h}, the gyroradius of each species sets a lower bound for the maximal distance of that species. Interestingly, the flute modes in several cases (particularly for the largest debris clouds with $R_m=150~d_i$) carry the lighter debris ions (data depicted by circles) farther out than the heavier species (data depicted by triangles).

\section{Estimate of 3D Effects}

The hybrid simulation methodology can be applied in 3D as well as 2D \cite{le:2019}. Nevertheless, a thorough systematic scan in 3D over the parameters we covered in our 2D scan would be prohibitively costly. While we expect a 3D explosion to be qualitatively similar to one in 2D, it would differ in certain quantitative details. We summarize here a few expected differences based on simple geometric arguments. First, Eq.~\ref{eq:rest} implies a different basic scaling depending on the dimensionality $D$, also reflected in Eq.~\ref{eq:deltamax}. We expect, however, that the fitting parameters $c_0$ and $c_1$ will be of similar magnitude in 2D and 3D based on the physical arguments that relate $c_0$ to a small gyroradius correction and $c_1$ to magnetic tension associated with nearly circular magnetic flux tubes. Second, there will be differences in the fraction of free-streaming debris ions between 2D and 3D caused by differences between the cylindrical and spherical geometries. In an initial 3D cloud of spherically expanding debris, the fraction of debris that can escape through a solid angle subtended by the critical angle $\theta_\ast$ \cite{bashurin:1983} is $F\sim[1-\cos(\theta_\ast)]\sim\theta_\ast^2/2$.  (Recall that it grows linearly in 2D as $F\sim2\theta_\ast/\pi$.) Therefore, the fitting function $\sigma(x)$ in Eq.~\ref{eq:f90} should be replaced by one that scales roughly as $x^2$ for small $x$. Likewise, in a 3D spherical expansion, a fraction $C_{90}=0.1$ of the initial debris has $|v_x|>0.9V_d$ (as opposed to $C_{90}\sim0.29$ in a 2D cylindrical expansion), and $C_{90}=0.1$ should be used in Eq.~\ref{eq:f90}.

An additional dimensional effect is that the transport caused by flute modes is exaggerated in 2D simulations. In a 2D simulation with magnetic field perpendicular to the plane, the magnetic flux tubes are perfectly straight and parallel, and thus they easily interchange positions. In a 3D astrophysical explosion, the diamagnetic cavity bends the field lines and produces a shear of the magnetic field. This tends to inhibit the growth of interchange modes. In addition, because the density of the expanding shell of debris drops roughly as $1/r^2$ in 3D (rather than $1/r$ in 2D), the density gradients that drive the flute modes can be weaker in 3D than in 2D for an explosion with the same values of equal mass radius $R_m$ and expansion Alfven Mach number $M_A$. Over the range of hypothetical parameters we considered, our 2D simulations give a rough upper limit on transport by the flute modes to at most a few debris gyroradii. Initial 3D Hybrid-VPIC simulations (not presented here) suggest this effect is weaker in 3D.

\section{Summary and Discussion}

We re-visited 2D hybrid (kinetic ion/fluid electron) simulation of astrophysical explosions, in which a rapidly expanding cloud of ionized debris couples via collisionless processes to an ambient magnetized background plasma. We studied simplified geometries with a uniform background magnetic field and plasma. By covering a much wider parameter space than was previously feasible, the simulation results confirm and elucidate aspects of prior studies of debris-ambient coupling \cite{golubev:1978, winske:2007}. This allowed us to formulate relatively simple scalings that highlight the basic physical mechanisms for debris-ambient coupling. The scalings interpolate across the range of parameters of typical space and laboratory \cite{heuer:2018} experiments. Real space and astrophysical systems will, of course, have more complex geometries including background gradients \cite{brecht:2009}. In these cases, the nominal plasma parameters vary across the explosion, and the processes that couple the debris to the ambient plasma may differ between regions of the explosion.

Our focus here was to nail down basic aspects of explosion phenomena, particularly the maximal size of the debris cloud and the fraction of the debris that escapes by streaming parallel to the magnetic field. The results qualitatively agree with an estimate based on energy partition \cite{clark:2013} up to semi-empirical fitting parameters. The simple model holds well when the debris ion gyroradius is small compared to the diamagnetic cavity. In the opposite case when the gyroradius is large, the ions ``decouple'' \cite{hewett:2011}, and the single-particle trajectories set a lower bound on the distance the debris ions travel. By considering cases with two ion species \cite{clark:2013}, with a minority species several times heavier than the other, we found that both regimes may co-exist when the gyroradius of the minority species is relatively large while the main debris species has a small gyroraradius. In addition, we studied debris ions that stream out from the cavity roughly parallel to the magnetic field. Semi-empirical scalings were again found, this time to estimate the fraction ($F_{90}$) of the debris that escapes from the explosion region at nearly ($>.90$ times) the initial explosion speed along the magnetic field. We also considered transport by flute modes, which occurs over time scales longer than could be explored previously \cite{winske:2007}. The 2D simulations give an upper bound of debris transport by the flute modes to a few debris gyroradii. 

The ion dynamical processes studied here have been the subject of experiments over the last several decades \cite{dimonte:1991,zakharov:2006,collette:2010, winske:2019} and more recently at UCLA in the Large Plasma Device (LAPD) device \cite{schaeffer:2015}. The experiments include a rich variety of additional plasma physics processes, including shocks \cite{schaeffer:2017} and electromagnetic ion streaming instabilities \cite{weidl:2016, heuer:2018}. Details of these processes are also amenable to 2D hybrid simulation studies  \cite{weidl:2019,heuer:2020} and will be considered in future work.

\appendix
\section{Simulation grid sizes and resolution}
The table below lists the grid sizes and resolution (normalized to the background ion inertial length $d_i$) for the hybrid simulations with a single debris ion species and with varying equal-mass radius $R_m$ and expansion Alfven-Mach number $M_A$ for background magnetic fields in the plane ($B_x$) and out of the plane ($B_y$). For each run that included an additional minority heavy ion species, we used a grid with $L_x=L_z=2000~d_i$ and a resolution of $\Delta x = 1~d_i$.

\begin{center}
 \begin{tabular}{|c c c c c c|} 
 \hline
 B      &$R_m$  &$M_A$  &$L_x/d_i$  &$L_z/d_i$  &$\Delta x/d_i$\\ [0.5ex] 
 \hline\hline
 $B_x$  &10     &0.5    &225    &225    &0.5\\ 
        &10     &1      &225    &225    &0.5\\
        &10     &2      &450    &225    &0.5\\ 
        &10     &5      &900    &225    &0.5\\ 
        &10     &10     &1350   &225    &0.5\\
        &40     &0.5    &270    &270    &0.5\\ 
        &40     &1      &540    &540    &0.5\\
        &40     &2      &1080   &540    &0.5\\ 
        &40     &5      &2160   &1080   &1\\ 
        &40     &10     &2160   &1080   &1\\
        &150    &0.5    &900    &900    &1\\ 
        &150    &1      &1800   &1350   &1\\
        &150    &2      &2700   &1800   &1\\ 
        &150    &5      &7200   &2000   &1\\ 
        &150    &10     &7200   &2000   &1\\ 
 \hline
 $B_y$  &10     &0.5    &135    &135    &0.5\\ 
        &10     &1      &135    &135    &0.5\\
        &10     &2      &270    &270    &0.5\\ 
        &10     &5      &270    &270    &0.5\\ 
        &10     &10     &270    &270    &0.5\\
        &40     &0.5    &270    &270    &0.5\\ 
        &40     &1      &540    &540    &0.5\\
        &40     &2      &540    &540    &0.5\\ 
        &40     &5      &1080   &1080   &1\\ 
        &40     &10     &1080   &1080   &1\\
        &150    &0.5    &900    &900    &1\\ 
        &150    &1      &900    &900    &1\\
        &150    &2      &1800   &1800   &1\\ 
        &150    &5      &2700   &2700   &1\\ 
        &150    &10     &2700   &2700   &1\\ 
 \hline
\end{tabular}
\end{center}

\acknowledgments
Research presented in this article was supported by the Laboratory Directed Research and Development program of Los Alamos National Laboratory under project 20200334ER and by the Defense Threat Reduction Agency under project DTRA1308134079. Computing resources were supplied by the LANL Institutional Computing program. Experimental data were not used, nor created for this research. The simulation data may be reproduced using the open-source VPIC code available at \url{github.com/lanl/vpic} with the hybrid modification described in the text and the numerical parameters outlined in the text and Appendix. The data sets used to produce the figures are available at \url{https://doi.org/10.5281/zenodo.5138065}.


%
%


%
%
%
%
%

\end{document}